\documentclass[aps,prx,reprint,groupedaddress,superscriptaddress,floatfix]{revtex4-2}

\usepackage[T1]{fontenc}
\usepackage{newtxtext}
\usepackage{newtxmath}

\usepackage{amsmath}
\usepackage{mathtools}
\usepackage{amsfonts}

\usepackage{color}
\usepackage{xcolor}
\usepackage[colorlinks,citecolor=blue,linkcolor=blue,urlcolor=blue]{hyperref}

\usepackage{graphicx}
\usepackage[all]{hypcap} 

\usepackage{makecell}

\begin{document}

\title{Calibrating the Classical Hardness of the Quantum Approximate Optimization Algorithm}

\author{Maxime Dupont}
\email[Corresponding author: ]{mdupont@rigetti.com}
\affiliation{Department of Physics, University of California, Berkeley, California 94720, USA}
\affiliation{Materials Sciences Division, Lawrence Berkeley National Laboratory, Berkeley, California 94720, USA}
\affiliation{Rigetti Computing, 775 Heinz Avenue, Berkeley, California 94710, USA}

\author{Nicolas Didier}
\affiliation{Rigetti Computing, 775 Heinz Avenue, Berkeley, California 94710, USA}

\author{Mark J. Hodson}
\affiliation{Rigetti Computing, 775 Heinz Avenue, Berkeley, California 94710, USA}

\author{Joel E. Moore}
\affiliation{Department of Physics, University of California, Berkeley, California 94720, USA}
\affiliation{Materials Sciences Division, Lawrence Berkeley National Laboratory, Berkeley, California 94720, USA}

\author{Matthew J. Reagor}
\affiliation{Rigetti Computing, 775 Heinz Avenue, Berkeley, California 94710, USA}

\begin{abstract}
    Trading fidelity for scale enables approximate classical simulators such as matrix product states (MPS) to run quantum circuits beyond exact methods. A control parameter, the so-called bond dimension $\chi$ for MPS, governs the allocated computational resources and the output fidelity. Here, we characterize the fidelity for the quantum approximate optimization algorithm by the expectation value of the cost function it seeks to minimize and find that it follows a scaling law $\mathcal{F}\bigl(\ln\chi\bigr/N\bigr)$ with $N$ the number of qubits. With $\ln\chi$ amounting to the entanglement that an MPS can encode, we show that the relevant variable for investigating the fidelity is the entanglement per qubit. Importantly, our results calibrate the classical computational power required to achieve the desired fidelity and benchmark the performance of quantum hardware in a realistic setup. For instance, we quantify the hardness of performing better classically than a noisy superconducting quantum processor by readily matching its output to the scaling function. Moreover, we relate the global fidelity to that of individual operations and establish its relationship with $\chi$ and $N$. We sharpen the requirements for noisy quantum computers to outperform classical techniques at running a quantum optimization algorithm in speed, size, and fidelity.
\end{abstract}

\maketitle

\section{Introduction}

Quantum advantage is achieved when a quantum computer outpaces a classical one at doing a specific task in speed, size, or quality~\cite{Preskill2012,Harrow2017,Boixo2018,Arute2019,Zhong2020,Madsen2022}. Whether a quantum processor succeeds is mainly a binary question. It can be hard to appreciate for noisy intermediate-scale quantum (NISQ) devices due to their small number of qubits and the presence of inherent noise, which limit the success rate of quantum algorithms~\cite{Preskill2018}. For instance, Google's ``Sycamore'' experiment sampling from a random circuit has been believed to be out of reach for classical hardware~\cite{Arute2019}. However, further development in tensor-network techniques has shown that the corresponding quantum circuits could be executed classically with better performance in a reasonable time~\cite{Huang2020,Feng2022}.

In order to run quantum circuits classically, a simulator is required. Exact tools include, e.g., state vectors, Feynman paths~\cite{Bernstein1997}, and tensor networks~\cite{Bridgeman2017} with some handling qubits in a superior manner over circuit depth and vice versa. Approximate simulators make the running 
of large-scale circuits feasible beyond exact ones. This comes at a price of lower fidelity in the execution, governed nontrivially by a control parameter $\chi$. These methods are mainly based on tensor networks such as matrix product states~\cite{Vidal2004}, projected entangled-pair states~\cite{Verstraete2004,Verstraete2006}, tree-tensor networks~\cite{Shi2006}, or the multiscale entanglement-renormalization ansatz~\cite{Vidal2008}.

\begin{figure}[b]
    \includegraphics[width=1\columnwidth]{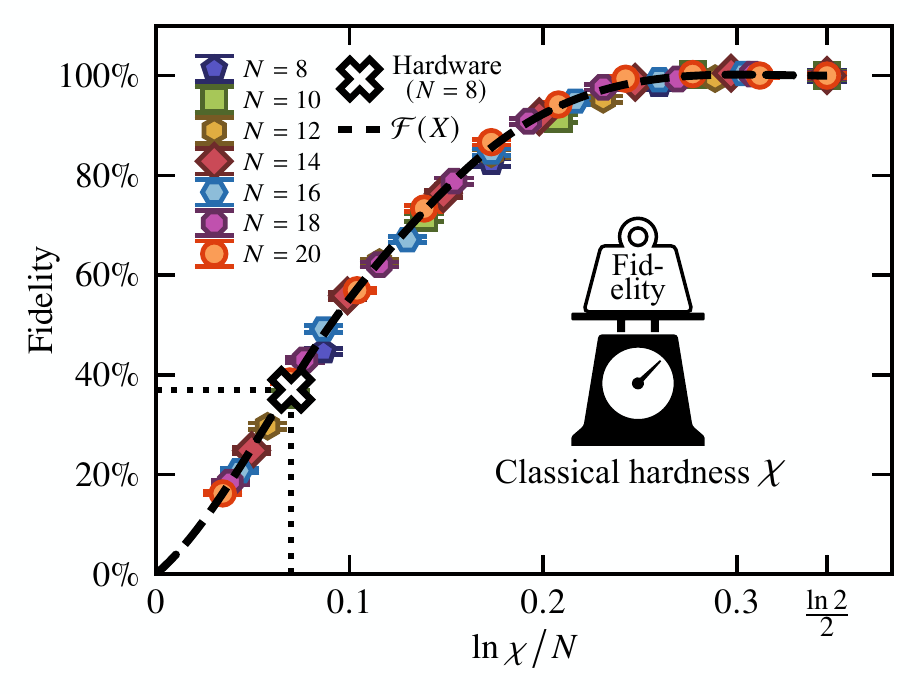} 
    \caption{A summary of the main results of this work. We perform MPS simulations with bond dimension $\chi$ of the QAOA for the Max-Cut problem on $N$ qubits. The best output cost that the algorithm seeks to minimize is used for defining the ``fidelity'' as the ratio $C\bigl(N,\chi\bigr)\bigr/C\bigl(N,\chi_\textrm{exact}\bigr)$, where $\chi_\textrm{exact}=2^{N/2}$ is the bond dimension required for exact calculations. Each data point corresponds to the statistical average of $100$ QAOA instances with one layer on random $3$-regular graphs. Plotting the data as a function of $X=\ln\chi/N$ makes them collapse onto a single curve $Y=\mathcal{F}(X)$, thus revealing a scaling relation. For instance, by matching the output of noisy quantum hardware, $Y=37(3)\%$, on $N=8$ qubits to the scaling function, one obtains a measure of the classical hardness $\chi=\textrm{exp}\bigl[\mathcal{F}^{-1}(Y)N\bigr]\approx 2$ of performing better than the quantum processor.}
    \label{fig:main_scaling}
\end{figure}

Matrix product states (MPSs)~\cite{Vidal2004} are perhaps best established, due to their simplicity and the extensive literature (for a review, see, e.g., Refs.~\cite{Schollwock2011,Orus2014}). Furthermore, we have a good understanding of how $\chi$, known as the bond dimension, affects the properties of a quantum state represented as a MPS. It amounts to the quantum entanglement that can be encoded in the system and it relates to the entanglement entropy $S\sim\ln\chi$. For $N$ qubits, the memory usage of MPSs scales as $O(N\chi^2)$ and the time complexity of individual computational operations scales at most as $O(\chi^3)$. Depending on the classical resources at hand, one can hope to perform calculations up to $\chi\approx 10^2-10^4$ (or effectively more when incorporating the symmetries of a circuit). For instance, in the context of quantum computing, MPSs have been used to simulate Google's ``Sycamore'' circuits~\cite{Zhou2020}, boson sampling~\cite{Oh2021}, Shor's algorithm~\cite{Shor1994,Wang2017,Dang2019}, and the quantum approximate optimization algorithm (QAOA)~\cite{Patti2021}.

The QAOA~\cite{Farhi2014,Farhi2014b,Farhi2016} is a quantum optimization algorithm seeking to solve quadratic unconstrained binary-optimization problems~\cite{Kochenberger2014} arising in finance, logistics, machine learning, and basic science, to cite but a few. These problems are often NP hard and there is no efficient classical algorithm to address them. For that reason, the QAOA is subject to intensive research~\cite{Farhi2014,Farhi2014b,Farhi2016,Wecker2016,Guerreschi2017,Otterbach2017,Jiang2017,Verdon2017,Lloyd2018,Crooks2018,Qiang2018,Wang2018,Lechner2018,Anschuetz2018,Sundar2019,Hastings2019,Hadfield2019,Guerreschi2019,Lykov2020,Dalzell2020,Bravyi2020,ZhouLeo2020,Wiersema2020,Wierichs2020,Pagano2020,Bengtsson2020,Wang2020,Willsch2020,Abrams2020,Patti2021,Medvidovic2021,Kremenetski2021,Barak2021,Fitzek2021,Juneseo2021,Dumitrescu2018,Herrman2021,Akshay2021,Harrigan2021,DiezValle2022,Zhu2022,Sohaib2022,Ebadi2022,Stollenwerk2022,Amaro2022,Lotshaw2022,GonzalezGarcia2022,Weidenfeller2022,Chen2022,Santra2022}.

Our main contribution is to readily evaluate the classical computational power required to obtain the desired QAOA output and thus provide a practical benchmark for NISQ devices to outperform approximate simulators at running circuits in both size and fidelity. Precisely, we investigate the performance of MPS simulations of QAOA circuits versus $N$ and $\chi$ for the paradigmatic Max-Cut problem. We reveal that analyzing the average output cost as a function of a single variable $\ln\chi/N$, which can be interpreted as the entanglement per degree of freedom, leads to data collapse onto a single curve, unveiling a scaling relation (see Fig.~\ref{fig:main_scaling}). This central result calibrates the classical hardness of obtaining the desired output fidelity, including that for quantum hardware. For instance, we benchmark the Rigetti Aspen-M-1 superconducting quantum chip on $N=8$ qubits by estimating its corresponding $\chi$. Besides, by relating the global fidelity to that of individual operations $f$, we determine how engineering efforts in increasing $N$ and improving $f$ translate into classical hardness. In addition to the average cost function, we show that a scaling relation involving $\chi$ and $N$ also emerges for the probability of finding the bit string that solves the Max-Cut problem, straightforwardly extending our results regarding the cost to this quantity. Finally, we find that deliberately computationally low-cost MPS QAOA simulations may speed up the optimization of finding good circuit parameters, adding to other strategies~\cite{Guerreschi2017,ZhouLeo2020,Wierichs2020}.

\section{Definitions and methods}

\subsection{Quantum approximate optimization algorithm}

A QAOA circuit for $N$ qubits with $p$ layers reads~\cite{Farhi2014,Farhi2014b,Farhi2016},
\begin{equation}
    \bigl\vert\boldsymbol{\beta},\boldsymbol{\gamma}\bigr\rangle=\left(\prod\nolimits_{\ell=1}^pU_{\beta_\ell}U_{\gamma_\ell}\right) H^{\otimes N}\vert 0\rangle^{\otimes N},
    \label{eq:qaoa_circuit}
\end{equation}
where $H$ is the Hadamard gate applied on the individual qubits. The parametrized unitaries read $U_{\beta_\ell}=\prod_j\exp(-i\beta_\ell X_j/2)$ and $U_{\gamma_\ell}=\exp(-i\gamma_\ell C/2)$. The cost-function operator related to the Max-Cut problem, $C=\sum_{\{i,j\}\in E}w_{ij}Z_iZ_j$~\footnote{Another custom definition of the Max-Cut cost function is $\tilde{C}=\sum_{\{i,j\}\in E}w_{ij}(1-Z_iZ_j)/2$, which directly counts the number of cuts that one seeks to maximize. Up to the constant $\sum_{\{i,j\}\in E}w_{ij}/2$ and a factor of $1/2$, maximizing $\tilde{C}$ is equivalent to minimizing $C=\sum_{\{i,j\}\in E}w_{ij}Z_iZ_j$.}, is defined for a graph $G=(V,E)$ with edges $\{i,j\}\in E$ carrying a weight $w_{ij}>0$ (a nonunit weight is often referred to as a weighted Max-Cut problem). $X_i$ and $Z_i$ are Pauli operators on qubit $i$. The goal is to classically optimize the parameters for the circuit output to minimize the cost function $C$. We carry out the minimization using the Broyden-Fletcher-Goldfarb-Shannon algorithm~\cite{BFGS1,BFGS2,BFGS3,BFGS4}. For each problem considered, we repeat the optimization procedure for approximately $10^3-10^4$ random initializations of the parameters and only keep the best result. The initial $2p$ parameters are drawn from the uniform distributions $\gamma\in[0,2\pi]$ and $\beta\in[0,\pi]$. Although we do not make use of them, we note that strategies have been proposed and benchmarked for speeding up the optimization of the QAOA~\cite{Guerreschi2017,ZhouLeo2020,Wierichs2020}.

\subsection{Matrix product states}

A MPS represents a quantum state in a local form, where each degree of freedom is associated with a tensor $A$. The tensors are arranged linearly~\cite{Vidal2004,Schollwock2011,Orus2014}:
\begin{equation}
    \bigl\vert\boldsymbol{\beta},\boldsymbol{\gamma}\bigr\rangle=\sum\nolimits_{\{\boldsymbol{s}\}}A^{s_1}_{a_1}A^{s_2}_{a_1a_2}A^{s_3}_{a_2a_3}\cdots A^{s_N}_{a_{N-1}}\vert\boldsymbol{s}\rangle,
    \label{eq:mps_definition}
\end{equation}
where Einstein summation notation is used. The physical index $s_i$ has a dimension of two (with value $0$ or $1$ for a qubit) and $a_i$ is the bond index of dimension $\chi$. One-qubit gates---Hadamard and $U_{\beta_\ell}$ in Eq.~\eqref{eq:qaoa_circuit}---are straightforward to apply, since the operation only involves a contraction between the gate and the corresponding tensor $A$ of the qubit. Unless specified otherwise, we represent the unitary $U_{\gamma_\ell}$ as a matrix product operator and apply it onto the MPS~\footnote{Another strategy, which can be scaled to larger systems, would be to decompose the unitary $U_{\gamma_\ell}$ as a product of $\vert E\vert$ two-qubit gates, individually applied on the corresponding MPS tensors with the use of internal $\textsc{SWAP}$ gates to accommodate for the native linear topology of MPSs. This other strategy is used in App.~\ref{app:cost_scaling} and lead to similar results.}. We keep the MPS bond dimension to the desired $\chi$ value through singular-value decompositions after each QAOA layer by keeping only the largest $\chi$ singular values and dropping the rest (the smallest ones). For reference, exact simulations require $\chi\equiv\chi_\textrm{exact}=2^{N/2}$. The state $\vert\boldsymbol{\beta},\boldsymbol{\gamma}\rangle$ is normalized at the end once the optimal angles are found. This strategy (i) helps to keep the cost landscape a smooth function of the angles~\cite{Sreedhar2022} and (ii) matches experimental observations (in a sense that will become clear in the following).

\section{Results}

\begin{figure}[t]
    \includegraphics[width=1\columnwidth]{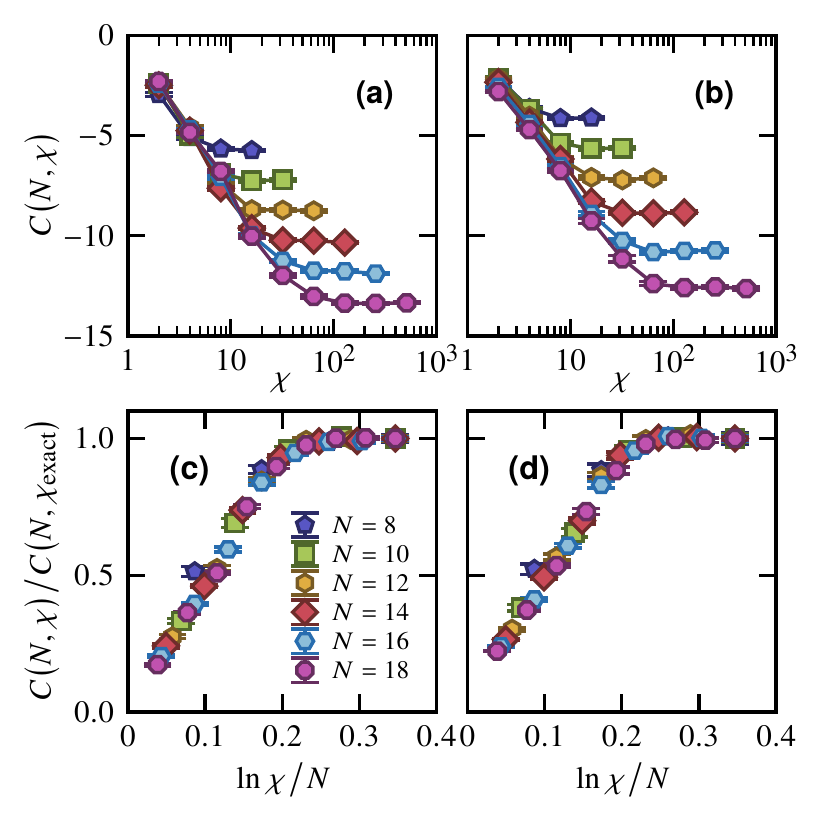} 
    \caption{Left column: $3$-regular graphs with unit weight and QAOA depth $p=2$. Right column: complete graphs with uniform random weight $w_{ij}\in[0,1]$ and QAOA depth $p=4$. Top row: the cost versus the bond dimension $\chi$ for different sizes $N$. Bottom row: the same data as the top row but with the $Y$ axis divided by the cost at $\chi_\textrm{exact}=2^{N/2}$ and the $X$ axis changed to $\ln\chi/N$. The data collapse onto a single curve $Y=\mathcal{F}(X)$, revealing a scaling law. Note the deviation for the smallest size $N=8$ from the scaling.}
    \label{fig:cost}
\end{figure}

We consider $3$-regular graphs with unit weight $w_{ij}=1$ together with QAOA depth $p=1$ (the case of Fig.~\ref{fig:main_scaling}) and $p=2$, as well as complete graphs with uniform random weights $w_{ij}\in[0,1]$ with QAOA depth $p=4$. For each of the three cases, we average the expectation value of the cost-function operator related to the Max-Cut problem over $100$ randomly generated graphs. We repeat the protocol versus $\chi$ for different sizes $N\leq 20$ and note the result $C(N,\chi)$. Due to the nature of the graphs considered, there is no obvious mapping to the native linear topology of a MPS and we randomly assign graph vertices to tensors. The data plotted in Figs.~\ref{fig:cost}(a) and (b) show that the cost decreases as the bond dimension increases. This suggests that, when rescaled by $C(N,\chi_\textrm{exact})$, the cost can serve as a proxy for the fidelity, i.e., a normalized number assessing the quality of the simulation, $1$ being perfect.

The exact cost $C(N,\chi_\textrm{exact})$ converges asymptotically with the circuit depth $p$ to the absolute minimum cost $C_\textrm{min}$. The absolute solution to the Max-Cut problem, for which there are $N_s$ bit strings $\boldsymbol{s}$ of cost $C_\textrm{min}$, can be written as $\sum_{\{\boldsymbol{s}\}}\textrm{e}^{i\phi_{\boldsymbol{s}}}\vert\boldsymbol{s}\rangle/\sqrt{N_s}$, where $\phi_{\boldsymbol{s}}$ a phase. If $N_s\sim O(1)$, the state has an exact MPS representation with finite $\chi$. For example, the complete graphs with randomized weights have two solutions ($N_s=2$): one bit string and its reverse obtained by the transformation $0\leftrightarrow 1$. This cat state, similar to a generalized Greenberger-Horne-Zeilinger (GHZ) state for $N$ qubits, has an exact MPS representation with $\chi=2$. The initial state $H^{\otimes N}\vert 0\rangle^{\otimes N}$ in the QAOA algorithm also has an exact MPS representation with $\chi=1$ (a product state). Nonetheless, simulating the QAOA in the restricted space of MPSs with bond dimension $\chi=2$ does not lead to the correct solution. This can be understood by studying the entanglement production and entanglement spreading in QAOA circuits, which need to accommodate volume-law entanglement $S\sim N$, requiring $\chi\sim\exp(N)$ in intermediate steps~\cite{Dupont2022}.

\subsection{Quantifying classical hardness}

\subsubsection{Scaling relation}
\label{sec:scaling_relation}

When plotting this effective fidelity as a function of $\ln\chi/N$, all the data points collapse onto a single curve $Y=\mathcal{F}(X)$ [see Fig.~\ref{fig:main_scaling} and Figs.~\ref{fig:cost}(c) and \ref{fig:cost}(d)]. This reveals a scaling relation of the form
\begin{equation}
    C\bigl(N,\chi\bigr)\Bigr/C\bigl(N,\chi_\textrm{exact}\bigr)=\mathcal{F}\Bigl(\ln\chi\bigr/N\Bigr),
    \label{eq:scaling}
\end{equation}
which is valid for all values of $\chi$ and $N$ considered, and supposedly any value. The relation holds for different graphs and different QAOA depths, suggesting universality---though the function $\mathcal{F}$ is different, as discussed in the following. The rescaled variable $\ln\chi/N$ can be interpreted as the entanglement per degree of freedom and is the relevant quantity to characterize the fidelity of QAOA circuits in MPS simulations (see also App.~\ref{app:entanglement_chi})~\cite{Dupont2022}.

Turning our attention to the scaling function $Y=\mathcal{F}(X)$, we find that in the limit of small $X$, it follows the functional form
\begin{equation}
    Y=\mathcal{F}\bigl(\ln\chi\bigr/N\to 0\bigr)\simeq A\bigl(\ln\chi\bigr/N\bigr)^\alpha,
    \label{eq:scaling_function_smallX}
\end{equation}
which is valid for $\ln\chi\bigr/N\lesssim 0.1$. The algebraic dependence is visible in Fig.~\ref{fig:misc}(a) for the three cases considered in this work. $A\sim O(1)$ and $\alpha\approx 1$ are parameters depending on the QAOA depth and graphs considered and which can be extracted by least-squares fitting.

Scaling laws involving $\chi$ typically arise in the context of MPSs representing quantum critical states in $1+1$ dimensions~\cite{Tagliacozzo2008,Pollmann2009}. In that case, a finite bond dimension induces a finite length scale $\xi\sim\textrm{poly}(\chi)$, the effect of which is analytically understood by conformal field theories. However, this is directly not applicable here, leaving the derivation of Eq.~\eqref{eq:scaling} as an open question. We believe that one might get further insights into the scaling relation by considering the evolution from the paramagnetic to the ferromagnetic ground state of the one-dimensional quantum Ising model. The corresponding circuit is analogous to the QAOA and achieves a similar purpose. In that case, we know that the critical point between the paramagnetic and ferromagnetic phases governs the properties of the final state with respect to the speed of the evolution process (known as Kibble-Zurek physics; for a review, e.g., Ref.~\cite{DelCampo2014}). As stated previously, plugging in a finite bond dimension in a static scenario modifies the critical properties by introducing a finite correlation length $\xi\sim\textrm{poly}(\chi)$~\cite{Tagliacozzo2008,Pollmann2009}. Therefore, we believe that investigating Kibble-Zurek physics in the context of finite-bond-dimension simulations would perhaps shed light on some of our findings.

Nevertheless, one can establish a rationale behind Eqs.~\eqref{eq:scaling} and~\eqref{eq:scaling_function_smallX}. As observed in Fig.~\ref{fig:cost}(a) and App.~\ref{app:cost_scaling}, for $\chi\ll\chi_\textrm{exact}$, the cost $C(N,\chi)$---non-normalized by data at $\chi_\textrm{exact}$---is independent of $N$ and shows a behavior consistent with $C(N,\chi)\sim\ln\chi$. In this regime, one can interpret the algorithm as having too little information to figure out how many qubits there are in the graph. It is analogous to the effect of a finite length scale $\chi$ in a critical system. As $\chi\to\chi_\textrm{exact}\sim\textrm{exp}(N)$, the cost saturates to an $N$-dependent value, as discussed in the following section~\ref{sec:exact_cost_normalization}. ,, the previous observation leads directly to $C(N,\chi)\sim N\times \mathcal{F}(\ln\chi/N)$ with $\mathcal{F}(X)\sim X$ at small $X$. This is consistent with the exponent $\alpha\approx 1$ found in Eq.~\eqref{eq:scaling_function_smallX} and the additional data of App.~\ref{app:cost_scaling}, from which we anticipate $\alpha=1$. In particular, we note that the fitted value $\alpha=1.10(3)>1$ in Fig.~\ref{fig:cost}(a) is unlikely to hold asymptotically, as it would imply that $C(N\to+\infty,\chi\ll\chi_\textrm{exact})\to 0$, while App.~\ref{app:cost_scaling} suggests otherwise, with the cost being constant at finite $\chi\ll\chi_\textrm{exact}$, independently of $N$. For complete graphs, whatever qubits are locally next to each other in the MPS are also connected in the underlying graphs. Hence, in the above picture, at fixed $\chi\ll\chi_\textrm{exact}$ but increasing $N$, the number of edges contributing to the total cost also increases, which is corroborated by the fact that $\alpha\lesssim 1$ in Eq.~\eqref{eq:scaling_function_smallX} and Fig.~\ref{fig:misc}(a) and by the data of Fig.~\ref{fig:cost}(b).

\subsubsection{Exact cost normalization}
\label{sec:exact_cost_normalization}

\begin{figure}[t]
    \includegraphics[width=1\columnwidth]{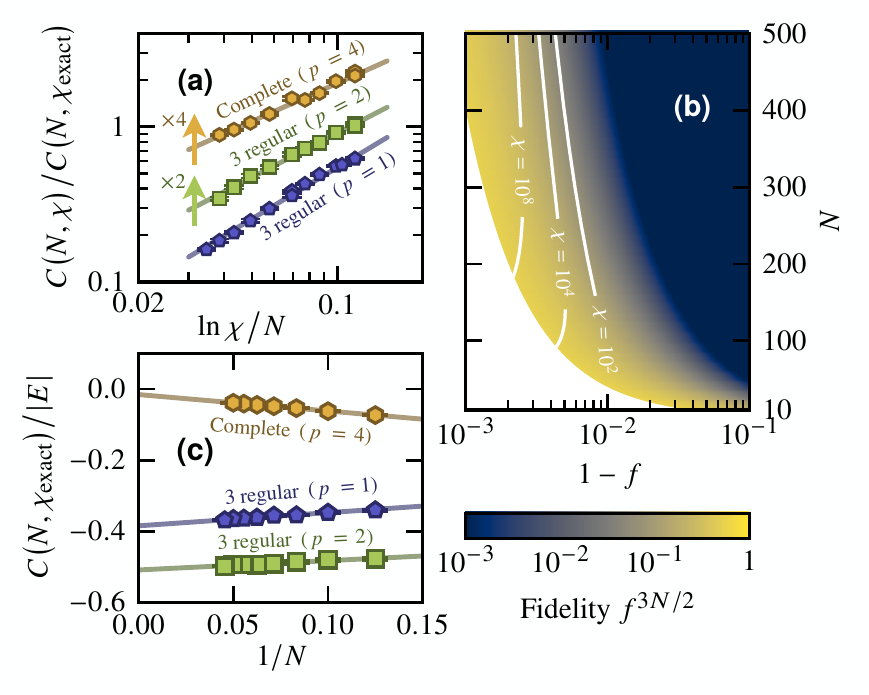} 
    \caption{(a) The scaling function $Y=\mathcal{F}(X\to 0)$ for unit-weight $3$-regular graphs with $p=1$ and $p=2$ (data shifted vertically by a factor of $2$ for readability) and complete graphs with uniform random weights $w_{ij}\in[0,1]$ and $p=4$ (data shifted vertically by a factor of $4$ for readability). The $N=8$ data are discarded due to their deviation to the scaling. The bold lines are a fit following Eq.~\eqref{eq:scaling_function_smallX} with parameters $A=6.9(6)$, $4.0(3)$, and $3.1(2)$, and $\alpha=1.10(3)$, $0.95(3)$, and $0.82(3)$, respectively. (b) The overall fidelity $f^K$ where $f$ is the average fidelity of individual operations and $K=3N/2$ is the number of operations. Using the parameters $A\approx 6.9$ and $\alpha\approx 1.10$ extracted from (a) for the case of unit-weight $3$-regular graphs with $p=1$, we plot iso-$\chi$ lines according to Eq.~\eqref{eq:fidelity}. (c) Extrapolation of the exact cost for arbitrary $N$ for the three cases considered. The straight lines are fits according to Eq.~\eqref{eq:cost_extrapolation}, with parameters $c_0=-0.385(2)$ and $c_1=0.37(3)$, $c_0=-0.509(3)$ and $c_1=0.26(4)$, and $c_0=-0.0157(4)$ and $c_1=-0.462(6)$, for unit-weight $3$-regular graphs with $p=1$ and $p=2$ and complete graphs with uniform random weights $w_{ij}\in[0,1]$ and $p=4$, respectively.}
    \label{fig:misc}
\end{figure}

The characterization of the fidelity of a QAOA circuit output by the expectation value of the cost that it seeks to minimize requires the exact cost $C(N,\chi_\textrm{exact})$ for normalization (see Eq.~\eqref{eq:scaling}). When considering a typical graph within an ensemble, the cost can be extrapolated from data at small $N$---not necessarily obtained with MPS simulations but simply any exact method. Indeed, we find that the average cost of the ensemble follows
\begin{equation}
    C\bigl(N,\chi_\textrm{exact}\bigr)\bigr/\vert E\vert\simeq c_0 + c_1\bigr/N + O\bigl(N^{-2}\bigr),
    \label{eq:cost_extrapolation}
\end{equation}
where $\vert E\vert$ is the total number of edges in the graph and $c_0$ and $c_1$ are fitting parameters. Such an extrapolation technique is customary to estimate the thermodynamic limit ($N\to+\infty$) ground-state energy density of quantum many-body Hamiltonians based on system sizes $N$ that can be simulated. We show in Fig.~\ref{fig:misc}(c) that Eq.~\eqref{eq:cost_extrapolation} works well for the three cases considered. Note that extrapolation of the exact cost $C(N,\chi_\textrm{exact})$ does not provide information regarding the quantum state $\vert\boldsymbol{\beta},\boldsymbol{\gamma}\rangle$ of Eq.~\eqref{eq:qaoa_circuit} for larger $N$ and thus does not solve the original Max-Cut problem of interest.

In perfect simulations, the quality of an output is measured by the approximation ratio $r=C(N,\chi_\textrm{exact})/C_\textrm{min}(N)$, where $C_\textrm{min}(N)$ is the absolute minimum cost. In the limit $p\to+\infty$, the definition of the fidelity of Eq.~\eqref{eq:scaling} is the same as $r$. At finite $p$, making the substitution $C(N,\chi_\textrm{exact})\to C_\textrm{min}(N)$ in Eq.~\eqref{eq:scaling} supposes that the two are related by a $N$-independent factor for the scaling relation to remain valid. This is found not to be the case~\cite{Crooks2018,ZhouLeo2020}, which we also confirm in App.~\ref{app:exact_vs_min}, although the small sizes accessible may bias the observation with respect to the large size limit.

\subsubsection{Extracting classical hardness}

Fig.~\ref{fig:main_scaling} and Figs.~\ref{fig:cost}(c) and \ref{fig:cost}(d) are calibration curves quantifying the classical computational power required for MPSs to achieve the desired fidelity $Y$ for a typical QAOA circuit. By inverting the scaling function $X=\mathcal{F}^{-1}(Y)$, one finds that the corresponding bond dimension for $N$ qubits is
\begin{equation}
    \chi(Y)=\exp\left[\mathcal{F}^{-1}(Y)N\right],
    \label{eq:inverse_scaling}
\end{equation}
which provides a direct measure of the classical hardness of running the circuit. In addition, the relationship shows that the difficulty increases exponentially with $N$. With quantum hardware heading toward $N\sim 10^3$ qubits in the coming years, a fidelity as low as $10\%$ for the QAOA on $3$-regular graphs with $p=1$ would require $\chi\sim 2\times 10^{9}$ (see Fig.~\ref{fig:main_scaling}), several order of magnitudes beyond the reach of classical computers. For comparison, as of today, the largest reported exact simulation of a QAOA circuit with $p=1$ on a $3$-regular graph has involved $210$ qubits~\cite{Lykov2020}. Besides, it has been argued that a quantum computer should contain at least $420$ flawless qubits for the QAOA to show a quantum advantage over classical algorithms solving the same class of problems~\cite{Dalzell2020} (see also Ref.~\cite{Guerreschi2019}).

\subsection{Benchmarking quantum hardware}

One can use Eq.~\eqref{eq:inverse_scaling} to benchmark a quantum device by extracting its corresponding $\chi$ on a calibration curve $Y=\mathcal{F}(X)$. We perform the experiment on the Rigetti Aspen-M-1 superconducting quantum chip. We consider the eight-vertices $3$-regular graph of Fig.~\ref{fig:exp_graph}(b) with unit weight together with QAOA depth $p=1$ (the case of Fig.~\ref{fig:main_scaling}). There are two parameters $\gamma_1$ and $\beta_1$, that we discretize on a two-dimensional grid. For each pair of parameters, we collect $1024$ output bit strings, from which we compute the expectation value of the cost-function operator of the Max-Cut problem. Additional experimental details are available in App.~\ref{app:hardware}.

\begin{figure}[t]
    \includegraphics[width=0.9\columnwidth]{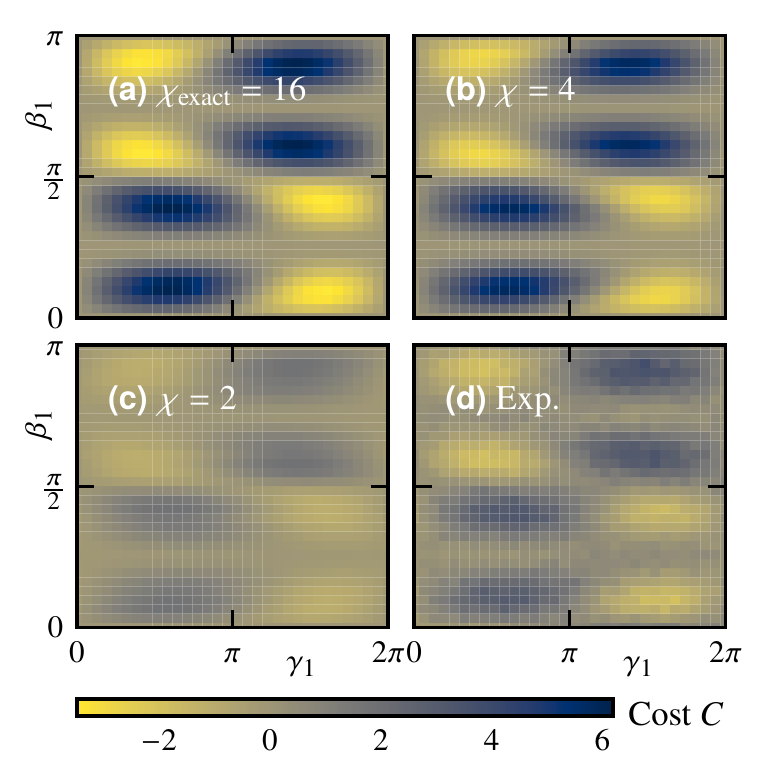} 
    \caption{The cost landscape for the $N=8$ unit-weight $3$-regular graph of Fig.~\ref{fig:exp_graph}(b) versus the two parameters $\gamma_1$ and $\beta_1$ of a one-layer QAOA circuit. (a), (b), and (c) Matrix-product-state simulations with $\chi\equiv\chi_\textrm{exact}=16$, $\chi=4$, and $\chi=2$, respectively. (d) The quantum processor output (Rigetti Aspen-M-1).}
    \label{fig:colormap}
\end{figure}

We plot the experimental result in Fig.~\ref{fig:colormap}(d). As a comparison, we perform MPS simulations with $\chi\equiv\chi_\textrm{exact}=16$, $\chi=4$, and $\chi=2$ [see Figs.~\ref{fig:colormap}(a), \ref{fig:colormap}(b), and \ref{fig:colormap}(c), respectively]. Hardware noise or small bond dimensions reduce the contrast without moving the position of the minimum very much, which has also been reported elsewhere~\cite{Qiang2018,Abrams2020,Willsch2020,Bengtsson2020,Pagano2020,Harrigan2021}. However, a low-fidelity QAOA does not solve the original problem. Indeed, a lessened contrast means a higher average cost, indicating that the output distribution of bit strings also follows this trend and that bit strings that minimize the cost function are less likely to appear. In App.~\ref{app:probab_bitstr}, we investigate the probability of finding bit strings with absolute minimum cost versus the bond dimension $\chi$ and the number of qubits $N$.

The minimum experimental cost in Fig.~\ref{fig:colormap}(d) is $C_\textrm{exp}=-1.9(1)$, associated with a fidelity of $46(3)\%$ with respect to the average exact cost $C(N=8,\chi_\textrm{exact})$. In Appendix~\ref{app:more_exp}, we repeat the experiment for all other other unit-weight $3$-regular graphs having $N=8$ vertices (there are five of them in total). Considering all the graphs, we find an average fidelity of $37(3)\%$, which we report in Fig.~\ref{fig:main_scaling}. We readily extract the corresponding average bond dimension $\chi\approx\textrm{exp}(0.07\times 8)\approx 2$, quantifying the classical hardness of doing at least as well using a MPS simulation. On a qualitative level, the MPS simulation with $\chi=2$ looks similar to the experimental data with a lower contrast compared to the exact simulation [see Figs.~\ref{fig:colormap}(c) and \ref{fig:colormap}(d), as well as Fig.~\ref{fig:colormap_all} in App.~\ref{app:more_exp}].

\subsection{Dissecting the fidelity}

It has been found that a good approximation of the overall fidelity of executing a quantum circuit follows the product of the fidelities of each individual operation (assuming no or very little correlations between errors)~\cite{Arute2019,Zhou2020}, including the expectation value of the QAOA cost function (see App. E in Ref.~\cite{Harrigan2021}). Hence, for a circuit comprised of $K$ operations with an average fidelity $f$, one finds a global fidelity decaying exponentially with $K$, i.e., $f^K$. For instance, a total of $16$ two-qubit gates are performed experimentally to generate Fig.~\ref{fig:colormap}(d) with an overall fidelity of $46(3)\%$. We estimate the effective fidelity as $f\approx 0.46^{1/16}\approx 95\%$, which is in line with the hardware specification, displaying an average fidelity of $96\%$ for the two-qubit gates (see App.~\ref{app:hardware}), assuming that fidelity of the two-qubit gates is the main source of errors.

When put in connection with the scaling relation of Eq.~\eqref{eq:scaling}, one can relate the number of qubits, the bond dimension, and $f$ together through $f^K\simeq\mathcal{F}(\ln\chi/N)$. Using the functional form of Eq.~\eqref{eq:scaling_function_smallX} for the scaling function, we obtain
\begin{equation}
    \chi\simeq\exp\left(A^{-1/\alpha}Nf^{K/\alpha}\right)~~~\textrm{for}~\ln\chi/N\lesssim 0.1.
    \label{eq:fidelity}
\end{equation}
For the case of unit-weight $3$-regular graphs with $p=1$, corresponding to Fig.~\ref{fig:main_scaling}, we obtain $A=6.9(6)$ and $\alpha=1.10(3)$, extracted from Fig.~\ref{fig:misc}(a). We use $K=3N/2$, corresponding to the number of edges in the graphs, and thus the minimum number of two-qubit gates required to be applied for the QAOA circuit, assuming all-to-all connectivity in the topology (this is the case in ion-trapped quantum computers) and that all other operations are perfectly executed. In Fig.~\ref{fig:misc}(b), we plot the overall fidelity as well as iso-$\chi$ lines versus $f$ and $N$. This shows that, for a few hundred qubits, even small improvements in $f$ can lead to a huge increase in the required bond dimension $\chi$ for MPS simulations to beat the hardware. For instance, for $N=300$ qubits in the regime $f\approx 99.4\%$, improving the error rate $1-f$ by a factor $1.5$, i.e., $f\approx 99.6\%$, makes the bond dimension increase by a factor of $100$. Fig.~\ref{fig:misc}(b) provides physicists and engineers with a quantitative idea of how incremental experimental efforts translate into classical hardness. We emphasize again that Eq.~\eqref{eq:fidelity} is valid in the regime $\ln\chi/N\lesssim 0.1$, which is the overall low-fidelity regime. Therefore, hundreds of qubits running a QAOA circuit with a global fidelity as low as approximately $10^{-2}-10^{-1}$ are out of reach for approximate MPS simulations. Per the central limit theorem, resolving such a low fidelity experimentally requires a number of shots scaling as its inverse square, which fits within the scope of current processor capabilities.

\section{Conclusions and outlook}

In summary, we address whether noisy qubits in the hundreds and thousands, which is where quantum devices are heading in the coming years, can have a computational advantage over a classical technique such as the MPS for the QAOA. The scaling relation shown in Fig.~\ref{fig:main_scaling} calibrates through $\chi$ how increasing the number of qubits and improving the fidelity of individual operations in a quantum computer makes it harder for classical simulators to beat it---and to eventually fail to do so.

All sources of noise are put on the same footing through an average fidelity parameter $f$. It would be interesting to perform noisy simulations to disentangle how specific errors (relaxation, dephasing, readout, etc.) modify the global fidelity of QAOA circuits, thus bringing the noise relation with $\chi$ one level closer to quantum hardware specifications.

We observe that for one-layer QAOA circuits, the main effect of a reduced bond dimension is to reduce the contrast of the cost landscape without moving the position of the minimum very much---this has also been reported elsewhere~\cite{Qiang2018,Abrams2020,Willsch2020,Bengtsson2020,Pagano2020,Harrigan2021}. If this holds for deeper QAOA circuits, running low-fidelity (i.e., low-cost) MPS simulations may reliably estimate the best angles $\{\boldsymbol{\gamma},\boldsymbol{\beta}\}$, which could then be used to speed up the optimization of larger-$\chi$ simulations. This would add to the other strategies that have been proposed~\cite{Guerreschi2017,ZhouLeo2020,Wierichs2020}.

A question left open is why a scaling relation emerges when considering the variable $\ln\chi/N$---or, equivalently, $\ln\chi/\ln\chi_\textrm{exact}$ or $S/N$ (see App.~\ref{app:entanglement_chi})---which, to the best of our knowledge, has not been reported elsewhere. It would be interesting to see whether such a scaling law appears in other contexts, such as variational~\cite{Cerezo2021} or quantum machine-learning algorithms~\cite{Schuld2015,Biamonte2017}, which ultimately seek to minimize a cost function. As reported in this work, this can be used to establish when and how noisy quantum computers will outperform classical techniques at running these circuits, both in size and fidelity. In addition, we report in App.~\ref{app:probab_bitstr} that there exists a scaling relation involving $\chi$ for the probability of finding the correct bit strings solving the Max-Cut problem with the QAOA. This suggests that the calibration strategy may apply to quantum algorithms with a unique bit string for an answer, such as Shor's and Grover's algorithms~\cite{Shor1994,Grover1996}---although it is computationally more challenging to evaluate the probability of finding a bit string than to evaluate the average of a cost function.

The calibration of classical hardness through the bond dimension is specific to MPSs, and it may be that other approximate classical simulators can perform better. For instance, while the topology of the graphs considered in this work does not clearly make one choice of tensor-network-based method better suited than another, it would be interesting to see how other techniques perform. Also, in the context of MPSs and for very specific graphs, a careful qubit ordering onto the linear MPS topology should lead to substantial improvements compared to a random assignment. Recently, shallow neural networks based on restricted Boltzmann machines~\cite{Hinton2002,Hinton2006,LeCun2015} have been used to simulate QAOA circuits~\cite{Medvidovic2021}, with promising results. Unlike our work based on MPSs, there is no easy way to estimate the classical difficulty of simulating the circuit for the desired fidelity and no understanding of what would control this---the entanglement per qubit with MPSs.

Note that outperforming classical techniques at running circuits is different than a quantum advantage. Indeed, there may exist classical algorithms solving---or attempting to solve---the problem of interest differently than by running a quantum circuit. For instance, even if solving the Max-Cut problem is NP hard, there exist efficient polynomial-time algorithms that find an approximate solution $r\%$ that is as good as the exact one with high probability. The Goemans-Williamson algorithm~\cite{Goemans1995} guarantees $r=87.8\%$ for general graphs, a value that can be improved in certain cases~\cite{Halperin2004}.

\begin{acknowledgments}
    We are indebted to one of the anonymous referees whose suggestions and insights led to the discussion at the end of Sec.~\ref{sec:scaling_relation} and App.~\ref{app:cost_scaling}. We acknowledge discussions with B. Evert, A. D. Hill, S. Jeffrey, S. L. Tomarken, and J. A. Valery. We also acknowledge discussions with S. Cohen, J.-S. Kim, and S. Stanwyck at Nvidia during the early stages of this work. J.E.M. was supported by the Quantum Science Center (QSC), a National Quantum Information Science Research Center of the U.S. Department of Energy (DOE), and a Simons Investigatorship. This research used the Lawrencium computational cluster resource provided by the IT Division at the Lawrence Berkeley National Laboratory (supported by the Director, Office of Science, Office of Basic Energy Sciences, of the U.S. Department of Energy under Award No. DE-AC02-05CH11231). This research also used resources of the National Energy Research Scientific Computing Center, a DOE Office of Science User Facility supported by the Office of Science of the U.S. Department of Energy under Contract No. DE-AC02-05CH11231 using NERSC Award No. DDR-ERCAP0022242. This research used resources of the Oak Ridge Leadership Computing Facility, which is a DOE Office of Science User Facility supported under Contract DE-AC05-00OR22725. The experimental results presented here are based upon work supported by the Defense Advanced Research Projects Agency (DARPA) under Agreement No. HR00112090058.
\end{acknowledgments}

\textit{Note added:} Recently, we became aware of Refs.~\cite{Sreedhar2022} and~\cite{Ayral2022} also investigating QAOA circuits with MPS.

\appendix

\section{Running on Rigetti Aspen-M-1}
\label{app:hardware}

\begin{figure}[t]
    \includegraphics[width=0.95\columnwidth]{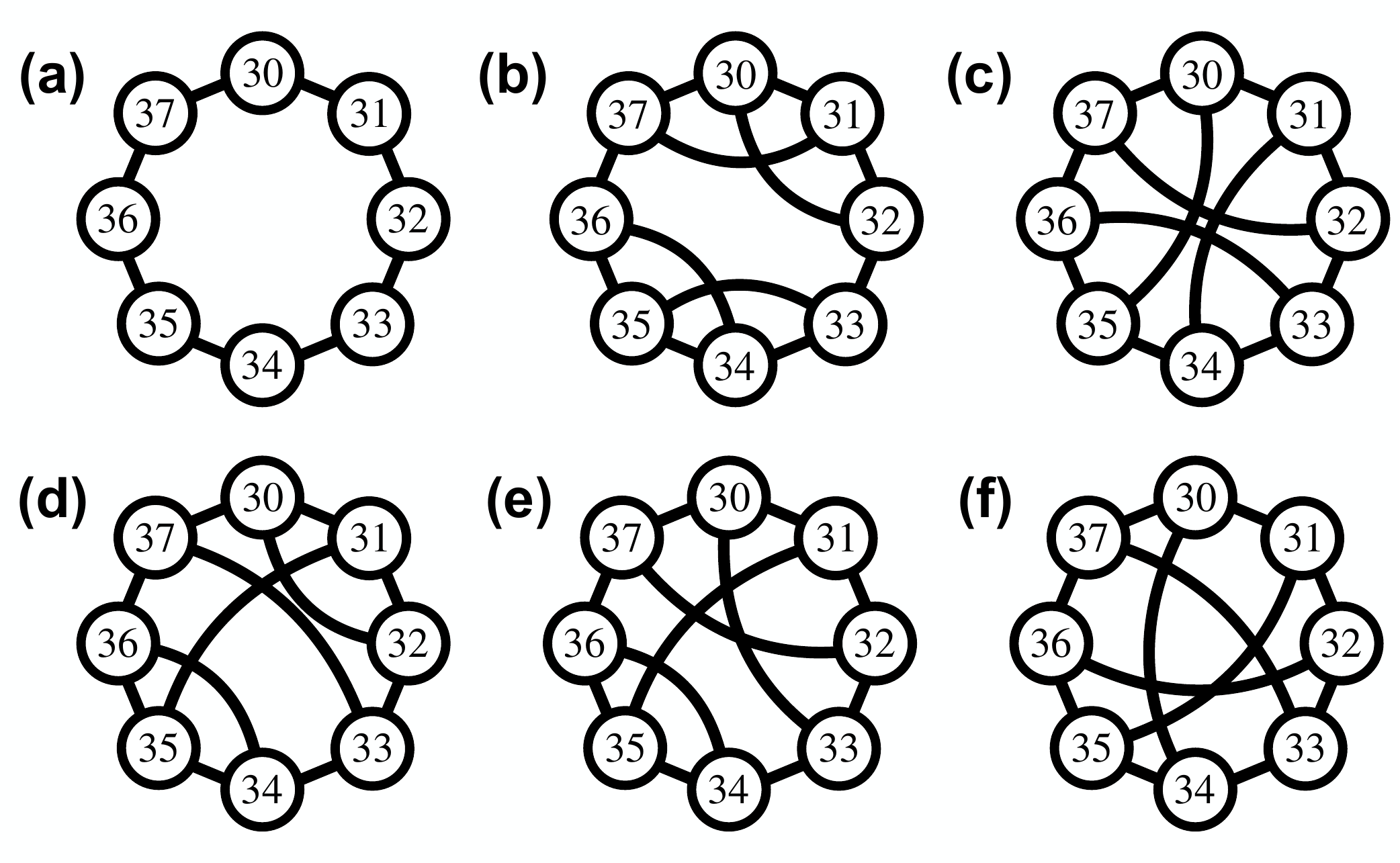} 
    \caption{(a) The native topology on Rigetti Aspen-M-1 for the eight qubits labeled $30$, $31$, $32$, $33$, $34$, $35$, $36$, and $37$ that we use for the experiment. (b)--(f) There exist five different  unit-weight $3$-regular graphs with $N=8$ vertices: (b) is considered in Fig.~\ref{fig:colormap} of the main text while (c)--(f) are considered in Fig.~\ref{fig:colormap_all}.}
    \label{fig:exp_graph}
\end{figure}

\subsection{Device specifications}

We use the eight qubits labeled $30$, $31$, $32$, $33$, $34$, $35$, $36$, and $37$ on Rigetti Aspen-M-1, which are connected in a ring topology [see Fig.~\ref{fig:exp_graph}(a)]. These qubits have an average relaxation time $T_1\simeq 29~\mu$s and dephasing time $T_2\simeq 38~\mu$s. One-qubit rotation about the $x$ axis $\textrm{Rx}(\mathbb{Z})=\exp(-i\mathbb{Z}\pi X/2)$ gates have an average fidelity of $99.6\%$ under single-qubit randomized benchmarking~\cite{Knill2008} and one-qubit rotation about the $z$ axis $\textrm{Rz}(\theta)=\exp(-i\theta Z/2)$ gates are virtual and thus error-free. The average readout fidelity is $96\%$. The native two-qubit gates are the controlled-phase $\textsc{CPHASE}(\theta)=\textrm{diag}(1, 1, 1, \textrm{e}^{i\theta})$ and $XY(\theta)=\exp[-i\theta (XX+YY)/2]$, both of which have an average fidelity of $96\%$, estimated by single two-qubit gate randomized benchmarking at $\theta=\pi$.

\subsection{QAOA gates with wative gates}

Implementing the QAOA with depth $p=1$ for $3$-regular graphs on the native ring topology of Fig.~\ref{fig:exp_graph}(a) requires the use of $\textsc{SWAP}$ gates. Up to one-qubit gates, it can be implemented as
\begin{equation}
    \textsc{SWAP}\sim XY(\pi)\textsc{CPHASE}(\pi).
    \label{eq:swap}
\end{equation}
The two-qubit gate $\exp(-i\gamma w_{ij}Z_iZ_j/2)$, where $\gamma$ is the QAOA angle and is $w_{ij}$ the weight of the edge between vertices $i$ and $j$, can be implemented using a single two-qubit gate (up to one-qubit gates):
\begin{equation}
    \exp\bigl(-i\gamma w_{ij}Z_iZ_j/2\bigr)\sim\textsc{CPHASE}\bigl(-2\gamma w_{ij}\bigr).
    \label{eq:expZZ}
\end{equation}

\section{Exact versus minimum cost}
\label{app:exact_vs_min}

\begin{figure}[!t]
    \includegraphics[width=1\columnwidth]{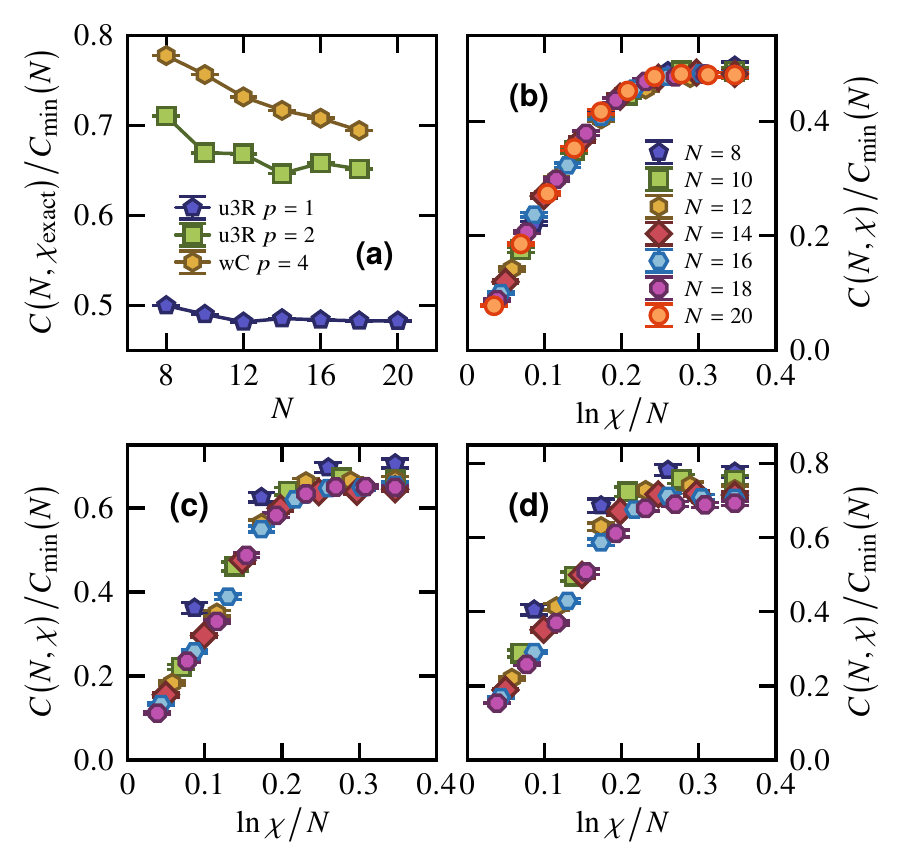} 
    \caption{(a) The approximation ratio $r=C(N,\chi_\textrm{exact})/C_\textrm{min}(N)$ versus the system size $N$ for unit-weight $3$-regular graphs with $p=1$ and $p=2$ and complete graphs with uniform random weights $w_{ij}\in[0,1]$ and $p=4$. (b)--(d) Data-collapse attempts according to Eq.~\eqref{eq:scaling}, substituting $C(N,\chi_\textrm{exact})$ by $C_\textrm{min}(N)$.}
    \label{fig:cmin_cexact}
\end{figure}

We investigate the relationship between the exact cost $C(N,\chi_\textrm{exact})$ for a fixed QAOA circuit and the absolute minimum $C_\textrm{min}(N)$ of the cost function that the QAOA seeks to minimize. The ratio $r=C(N,\chi_\textrm{exact})/C_\textrm{min}(N)$ asymptotically converges to $1$ with QAOA depth $p$.

We find that $r$ shows an $N$ dependence [which is especially visible for larger $p$ values; see Fig.~\ref{fig:cmin_cexact}(a)]. It was also observed in other works~\cite{Crooks2018,ZhouLeo2020}. Because exact calculations can only be performed on small systems, one cannot exclude the possibility that this dependence will disappear as $N\to+\infty$. With $C(N,\chi_\textrm{exact})$ and $C_\textrm{min}(N)$ related by an $N$-dependent constant, the scaling relation Eq.~\eqref{eq:scaling} does not hold when substituting $C(N,\chi_\textrm{exact})$ by $C_\textrm{min}(N)$. We attempt the data collapse anyway in Figs.~\ref{fig:cmin_cexact}(b)--\ref{fig:cmin_cexact}(d), which works reasonably well for the unit-weight $3$-regular graphs where $p=1$, as the corresponding $r=C(N,\chi_\textrm{exact})/C_\textrm{min}(N)$ curve is mostly flat versus $N$. If the approximation ratio $r$ was to flatten versus $N$ in a generic case for a fixed QAOA circuit as $N\to+\infty$, one could probably get good data collapse even for small system sizes by including corrections to the scaling~\cite{Beach2005,Wang2006}.

\section{Relation between the bond dimension and the entanglement}
\label{app:entanglement_chi}

\begin{figure}[!t]
    \includegraphics[width=1\columnwidth]{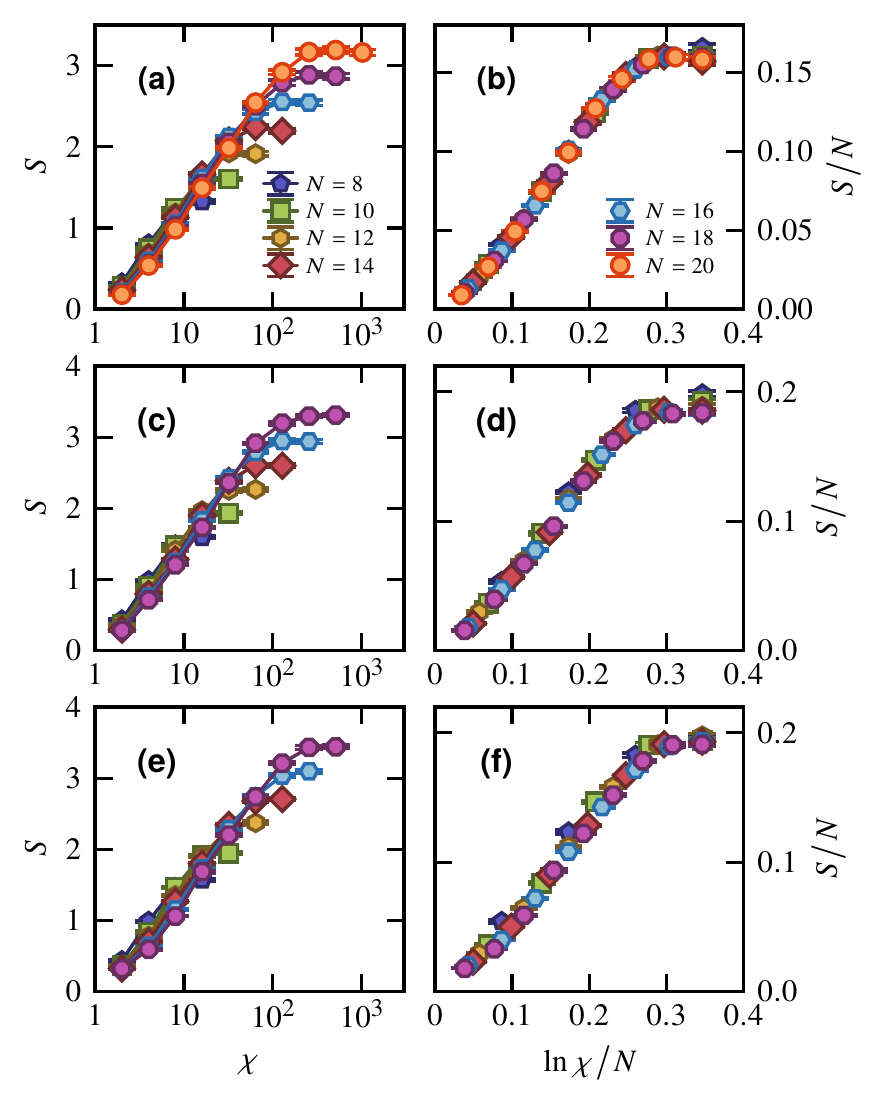} 
    \caption{Top row: $3$-regular graphs with unit weight and QAOA depth $p=1$. Middle row: $3$-regular graphs with unit weight and QAOA depth $p=2$. Bottom row: complete graphs with uniform random weight $w_{ij}\in[0,1]$ and QAOA depth $p=4$. Left column: the average bipartite Von Neumann entanglement entropy $S$ from a cut in the middle of the MPS versus the bond dimension $\chi$. Right column: the average bipartite Von Neumann entanglement entropy $S$ from a cut in the middle of the MPS rescaled by the system size $N$ versus the rescaled variable $\ln\chi/N$.}
    \label{fig:entanglement_chi}
\end{figure}

In pure states, the entanglement entropy puts a number on the degree of quantum entanglement between two subsets of qubits $A$ and $B$ of a system defined over $A\cup B$. The reduced density matrix of the subsystem $A$, $\rho_A=\mathrm{tr}_B\vert\boldsymbol{\beta},\boldsymbol{\gamma}\rangle\langle\boldsymbol{\beta},\boldsymbol{\gamma}\vert$ of the pure state $\vert\boldsymbol{\beta},\boldsymbol{\gamma}\rangle$ is used to compute the bipartite Von Neumann entanglement entropy between $A$ and $B$:
\begin{equation}
    S = -\mathrm{tr}\bigl(\rho_A\ln\rho_A\bigr)=-\sum\nolimits_k\lambda_k\ln\lambda_k,
    \label{eq:entanglement}
\end{equation}
where the $\lambda_k$ are the eigenvalues of $\rho_A$. Here, we compute $S$ by cutting the system in half with respect to the left- and right-hand sides of the linear MPS topology according to Eq.~\eqref{eq:mps_definition}. The setting is the same as in the main text and we consider the same three cases: unit-weight $3$-regular graphs, QAOA depth $p=1$ and $p=2$, as well as complete graphs with uniform random weights $w_{ij}\in[0,1]$ and QAOA depth $p=4$.

\begin{figure}[!t]
    \includegraphics[width=1\columnwidth]{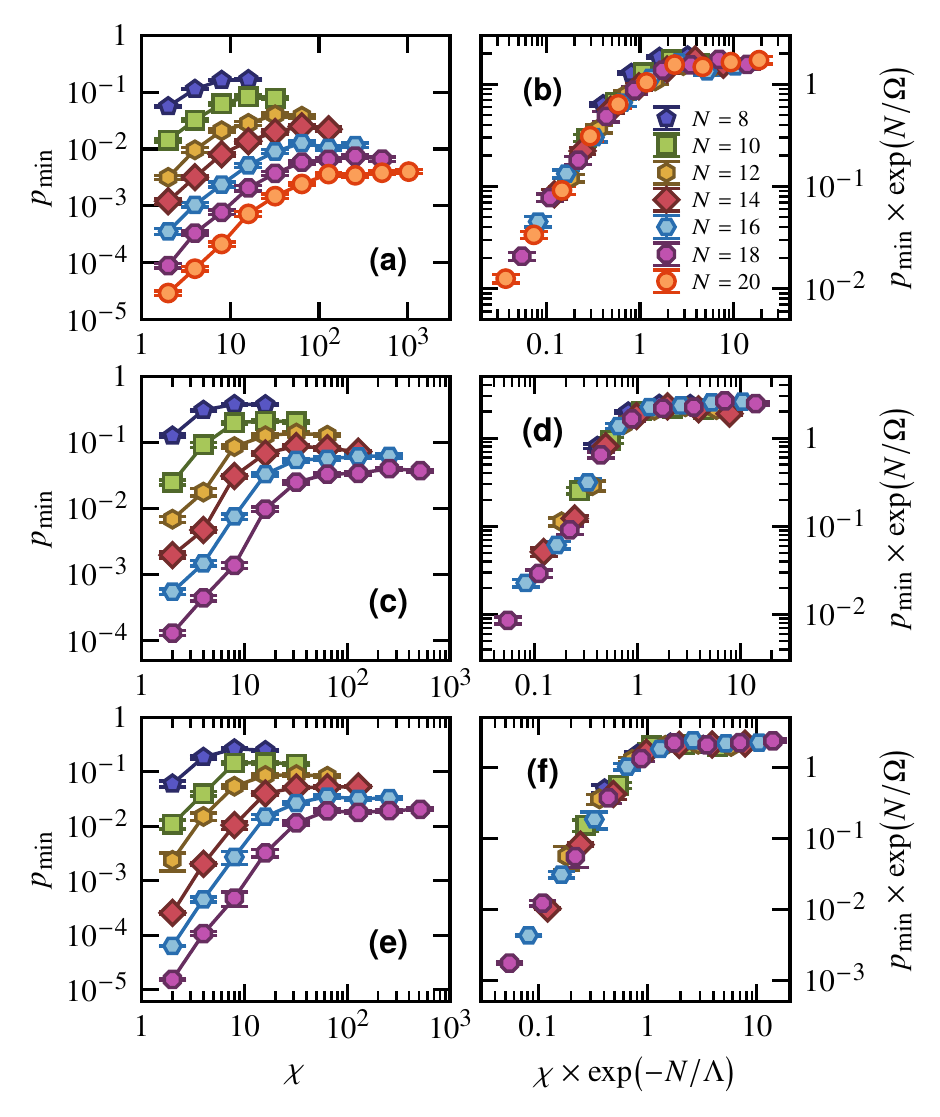} 
    \caption{Top row: $3$-regular graphs with unit-weight and QAOA depth $p=1$. Middle row: $3$-regular graphs with unit-weight and QAOA depth $p=2$. Bottom row: complete graphs with uniform random weight $w_{ij}\in[0,1]$ and QAOA depth $p=4$. Left column: the probability $p_\textrm{min}$ of finding a bit string minimizing the cost function of the Max-Cut problem versus the bond dimension $\chi$ for various system sizes $N$. Right column: the same data as the left column except that the $x$ axis has been rescaled by $\exp(-N/\Lambda)$ and the $y$ axis by $\exp(N/\Omega)$, where $\Lambda$ and $\Omega$ are two parameters such that the data collapse. We find $\Lambda\approx 5$ in the three cases and $\Omega\approx 3.3$, $4.3$, and $3.8$ in (b), (d), and (f), respectively.}
    \label{fig:probability_chi}
\end{figure}

In Figs.~\ref{fig:entanglement_chi}(a), \ref{fig:entanglement_chi}(c), and \ref{fig:entanglement_chi}(e), we plot the bipartite Von Neumann entanglement entropy and $S$ versus the bond dimension $\chi$. We verify that $S\sim\ln\chi$ until the entropy saturates for large $\chi$ to a volume-law value, i.e., $S\propto N$~\cite{Dupont2022}. In Figs.~\ref{fig:entanglement_chi}(b), \ref{fig:entanglement_chi}(d), and \ref{fig:entanglement_chi}(f), we display the entanglement per qubit versus the rescaled variable $\ln\chi/N$, which is the relevant variable of the scaling relation of Eq.~\eqref{eq:scaling}. We find good data collapse, except for small system sizes $N$, especially in the limit of small bond dimensions $\chi$. Discarding these data points, the collapse suggests the relation $S/N\simeq\mathcal{G}(\ln\chi/N)$, were $\mathcal{G}$ is a non-universal function with a linear behavior at small $\ln\chi/N$ (from which one recovers $S\sim\ln\chi$) before it saturates. This means that the variable $\ln\chi/N$ in Eq.~\eqref{eq:scaling} can indeed be interpreted as the entanglement per qubit $S/N$.

\section{Cost versus size and bond dimension}
\label{app:cost_scaling}

When considering the cost $C(N,\chi)$ without the need for rescaling by the cost of exact simulations, we can push to larger sizes with $\chi\ll\chi_\textrm{exact}$. Note that unlike in other sections of the paper, we do not use a matrix-product-operator representation of the quantum circuit but we apply one-qubit and two-qubit gates individually on the matrix product state representing the quantum state.

\begin{figure}[!t]
    \includegraphics[width=0.95\columnwidth]{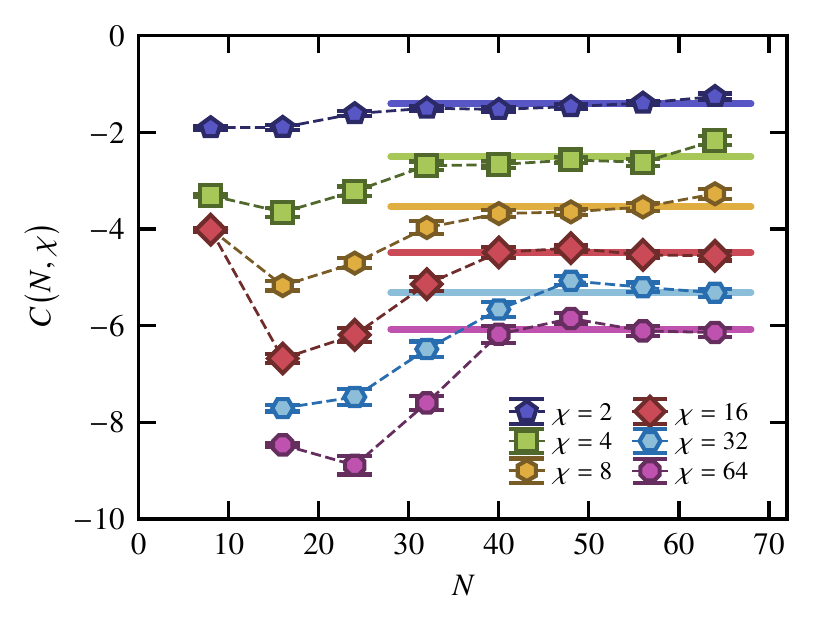} 
    \caption{The cost versus the size $N$ of $3$-regular graphs with a unit weight (QAOA depth $p=1$) for various values of the bond dimension $\chi$. Each data point corresponds to the statistical average of $100$ independent instances. The straight solid lines are a guide to the eye, highlighting that at fixed $\chi\ll\chi_\textrm{exact}$, the cost is roughly constant versus the size $N$. Note that unlike other plots in the paper, we do not use a matrix-product-operator representation of the quantum circuit but we apply one-qubit and two-qubit gates individually on the matrix product state representing the quantum state.}
    \label{fig:cost_scaling}
\end{figure}

We focus on $3$-regular graphs with a unit weight and a QAOA depth $p=1$. We show the data in Fig.~\ref{fig:cost_scaling} as a function of $N$ for various values of the bond dimension $\chi$. For $\chi\ll\chi_\textrm{exact}$, we find that the cost is independent of $N$ and saturates to a finite value that decreases as $\chi$ increases. At fixed $N$, the cost behaves as $C(N,\chi)\sim\ln\chi$. This behavior was also observed in Figs.~\ref{fig:cost}(a) and~\ref{fig:cost}(b) in different settings for smaller sizes. This data is valuable in Sec.~\ref{sec:scaling_relation} when establishing a rationale behind the emergence of the scaling function of Eq.~\eqref{eq:scaling}.

\begin{figure*}[t]
    \includegraphics[width=1.85\columnwidth]{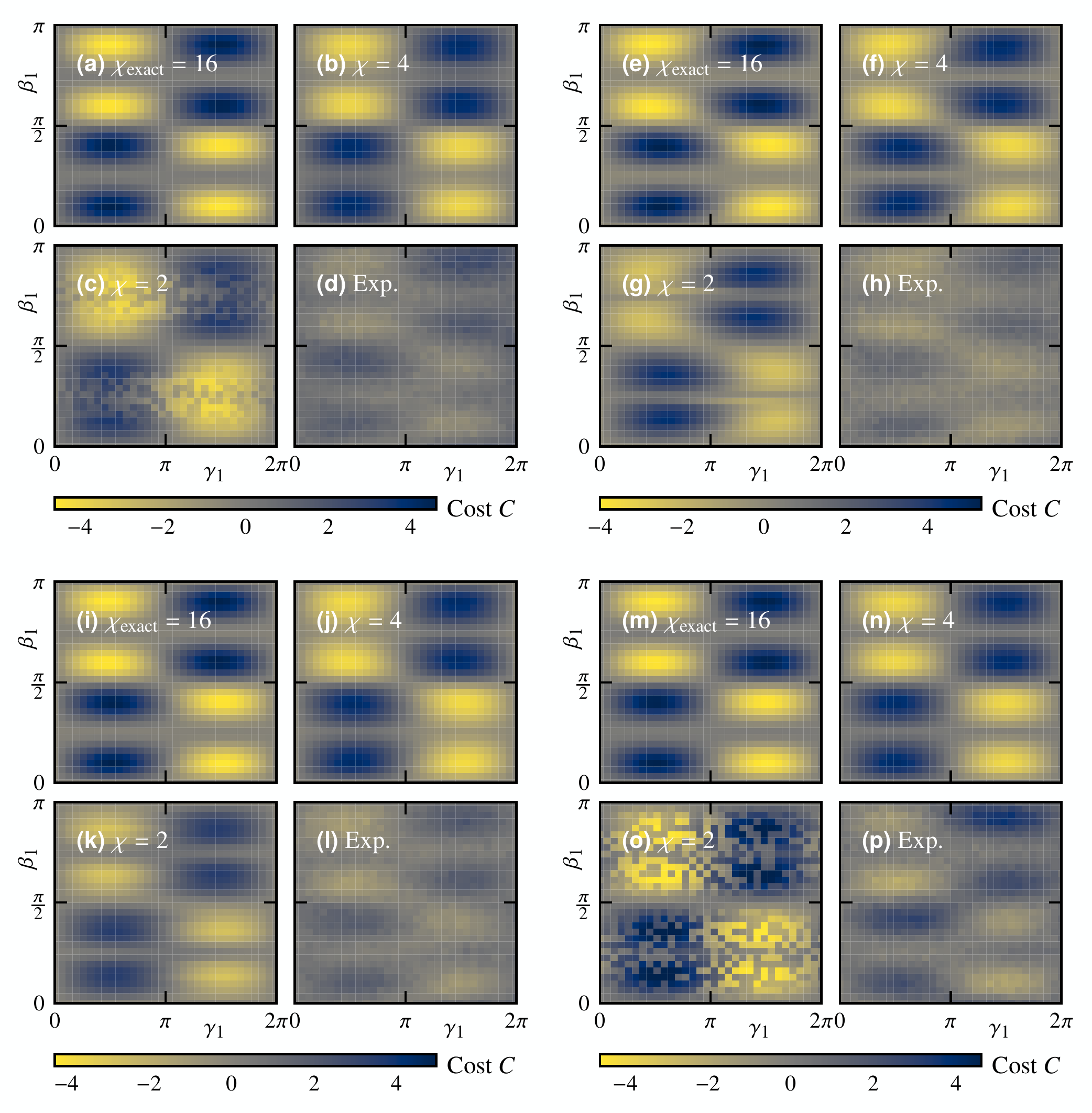} 
    \caption{The cost landscape for $N=8$ unit-weight $3$-regular graphs versus the two parameters $\gamma_1$ and $\beta_1$ of a one-layer QAOA circuit---a comparison of the matrix-product-state simulations with $\chi\equiv\chi_\textrm{exact}=16$, $\chi=4$, $\chi=2$ together with the quantum processor output (Rigetti Aspen-M-1). (a)--(d) $3$-regular graph of Fig.~\ref{fig:exp_graph}(f). (e)--(h) $3$-regular graph of Fig.~\ref{fig:exp_graph}(d). (i)--(l) $3$-regular graph of Fig.~\ref{fig:exp_graph}(e). (m)--(p) $3$-regular graph of Fig.~\ref{fig:exp_graph}(c).}
    \label{fig:colormap_all}
\end{figure*}

\section{Probability of finding the absolute solution to the max-cut problem}
\label{app:probab_bitstr}

Through a cost-function minimization, the QAOA seeks to output bit strings $\{\boldsymbol{s}\}$ with absolute minimum cost. These bit strings solve the Max-Cut problem of interest. The probability of finding these bitstrings is
\begin{equation}
    p_\textrm{min} = \sum\nolimits_{\{\boldsymbol{s}\}}\bigl\vert\langle\boldsymbol{s}\vert\boldsymbol{\beta},\boldsymbol{\gamma}\rangle\bigr\vert^2,
    \label{eq:pmin}
\end{equation}
where $\langle\boldsymbol{\beta},\boldsymbol{\gamma}\vert\boldsymbol{\beta},\boldsymbol{\gamma}\rangle=1$. For small enough systems, we can enumerate all bit srings to find the set $\{\boldsymbol{s}\}$. In Figs.~\ref{fig:probability_chi}(a), \ref{fig:probability_chi}(c), and \ref{fig:probability_chi}(e), we plot the probability versus the bond dimension $\chi$ for different system sizes, for the same three cases considered throughout this work: unit-weight $3$-regular graphs, QAOA depth $p=1$ and $p=2$, as well as complete graphs with uniform random weights $w_{ij}\in[0,1]$, QAOA depth $p=4$. We observe that the probability decreases with the system size and increases with the bond dimension.

We find that the data points collapse onto a single curve with the following scaling relation:
\begin{equation}
    p_\textrm{min}=\mathrm{e}^{-N/\Omega}\,\mathcal{H}\Bigl(\chi\mathrm{e}^{-N/\Lambda}\Bigr),
    \label{eq:pmin_scaling}
\end{equation}
where $\Omega$ and $\Lambda$ are parameters, and $\mathcal{H}$ is a scaling function, all of which are a priori non-universal. For $X\equiv\chi\mathrm{e}^{-N/\Lambda}\lesssim 1$, we observe an algebraic dependence for the scaling function $\mathcal{H}(X)\simeq DX^\delta$ with parameters $D$ and $\delta$ that can be extracted by least-squares fitting. For instance, for the case of unit-weight $3$-regular graphs with QAOA depth $p=1$, we find $D\approx 1.9$ and $\delta\approx 1.6$. This scaling shows that in its regime of validity, at fixed bond dimension $\chi$, the probability decreases exponentially with $N$. Similarly, to maintain a fixed probability, the bond dimension has to scale exponentially with the system size. By randomly sampling bitstrings, the probability of finding the solution $\propto 2^{-N}$ is also exponentially small with the system size. However, the QAOA makes the factor in the exponential more favorable. Although it is beyond the scope of this work, one may expect that additional QAOA layers in the circuit will also make it more favorable.

\section{Additional experiments}
\label{app:more_exp}

\begin{figure}[t]
    \centering
    \includegraphics[width=0.9\columnwidth]{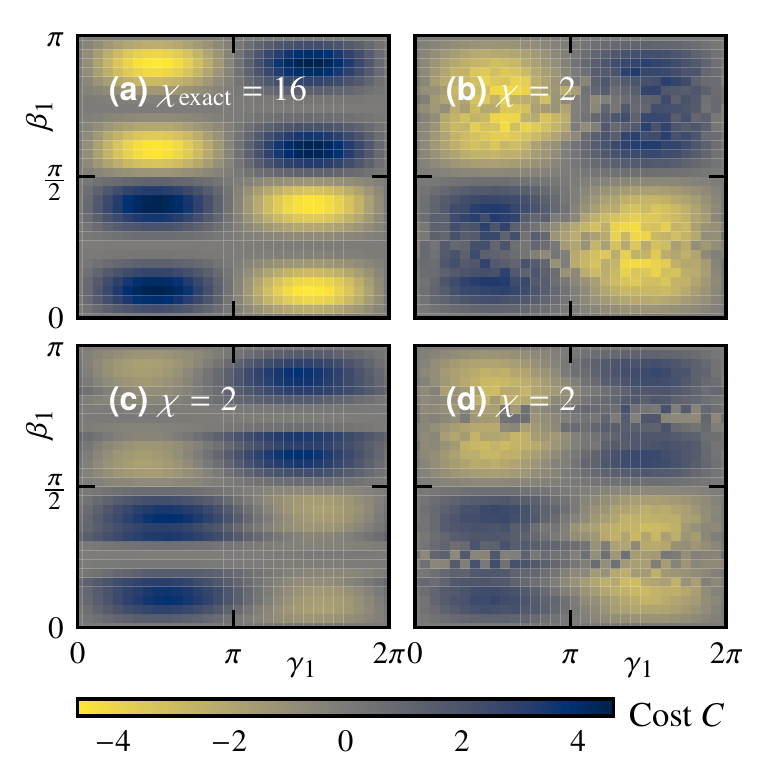} 
    \caption{The cost landscape for the $N=8$ unit-weight $3$-regular graph of Fig.~\ref{fig:exp_graph}(f) versus the two parameters $\gamma_1$ and $\beta_1$ of a one-layer QAOA circuit. (a) An exact simulation with $\chi=16$. (b)--(d) An approximate simulation with $\chi=2$. The mapping from the graph to the one-dimensional topology of the MPS is different in (b)--(d). One observes that the truncation to keep $\chi=2$ affects the landscape differently depending on the mapping.}
    \label{fig:colormap_unfolding}
\end{figure}

In addition to the $3$-regular graph of Fig.~\ref{fig:exp_graph}(b) discussed in the main text, we consider all the other unit-weight $3$-regular graphs with eight vertices, which are displayed in Figs.~\ref{fig:exp_graph}(c), \ref{fig:exp_graph}(d), \ref{fig:exp_graph}(e), and \ref{fig:exp_graph}(f). For a single-layer QAOA, the two parameters $\gamma_1$ and $\beta_1$ are discretized on a two-dimensional grid. For each pair of parameters, we collect $1024$ output bit strings from which we compute the expectation value of the cost-function operator of the Max-Cut problem.

We plot the experimental results in Fig.~\ref{fig:colormap_all} and compare with MPS simulations using $\chi\equiv\chi_\textrm{exact}=16$, $\chi=4$, and $\chi=2$. As observed previously, hardware noise reduces the contrast without changing the landscape too much otherwise. The discontinuous behavior in the landscape, observed in Figs.~\ref{fig:colormap_all}(c) and (o) is due to the choice of mapping from the $3$-regular graph to the one-dimensional topology of the MPS. It is verified in Fig.~\ref{fig:colormap_unfolding}, which considers different mappings of the same problem. The relevant numerical values related to the experiments are reported in Table~\ref{tab:numerical_data}. The fidelity associated with the graphs of Figs.~\ref{fig:exp_graph}(d), \ref{fig:exp_graph}(e), and \ref{fig:exp_graph}(f) is lower than for the graphs of Figs.~\ref{fig:exp_graph}(b) and \ref{fig:exp_graph}(c) due to the higher number of $\textsc{SWAP}$ gates required to map the graphs to the native topology.

\begin{table}[!b]
    \vspace{0.8em}
		\center
		\begin{ruledtabular}
			\begin{tabular}{c|cccccc}
				\thead{\textbf{Graph}} & \thead{\textbf{Exp.}\\\textbf{Min. Cost}} & \thead{\textbf{Exact}\\\textbf{Min. Cost}} & \thead{\textbf{Fidelity}} & \thead{\textbf{\#}\\$\textsc{SWAP}$} & \thead{\textbf{\# Layers}\\\textbf{2Q Gates}} & \thead{\textbf{$f$}}\\
				\hline\\[-0.8em]
				\makecell{Fig.~\ref{fig:exp_graph}(b)} & \makecell{$-1.9(1)$} & \makecell{$-3.484$} & \makecell{$46(3)\%$} & \makecell{$2$} & \makecell{$5$} & \makecell{$95.3(4)\%$}\\[0.3em]
				\makecell{Fig.~\ref{fig:exp_graph}(c)} & \makecell{$-2.4(2)$} & \makecell{$-4.612$} & \makecell{$57(3)\%$} & \makecell{$4$} & \makecell{$5$} & \makecell{$97.2(3)\%$}\\[0.3em]
				\makecell{Fig.~\ref{fig:exp_graph}(d)} & \makecell{$-1.0(1)$} & \makecell{$-4.005$} & \makecell{$24(3)\%$} & \makecell{$5$} & \makecell{$11$} & \makecell{$93.7(6)\%$}\\[0.3em]
				\makecell{Fig.~\ref{fig:exp_graph}(e)} & \makecell{$-1.3(1)$} & \makecell{$-4.285$} & \makecell{$32(3)\%$} & \makecell{$4$} & \makecell{$7$} & \makecell{$94.5(5)\%$}\\[0.3em]
				\makecell{Fig.~\ref{fig:exp_graph}(f)} & \makecell{$-1.1(1)$} & \makecell{$-4.612$} & \makecell{$26(3)\%$} & \makecell{$6$} & \makecell{$7$} & \makecell{$94.5(5)\%$}\\
				\hline\\[-0.8em]
				\makecell{\textbf{Average}} & \makecell{$-1.5(1)$} & \makecell{$-4.200$} & \makecell{$37(3)\%$} & \makecell{4.2} & \makecell{7} & \makecell{$95.0(5)\%$}
			\end{tabular}
		\end{ruledtabular}
	\caption{The relevant numerical data for the five unit-weight $3$-regular graphs with $N=8$ vertices that exist. The fidelity is computed with respect to the average exact minimum cost. The individual fidelity $f$ is computed from the global fidelity assuming a number of operations equal to the number of two-qubit gates applied in the circuit. This number equals $12$ (number of edges in a $3$-regular graph with $N=8$ vertices) plus twice the number of $\textsc{SWAP}$ gates according to Eqs.~\eqref{eq:expZZ} and~\eqref{eq:swap}.}
	\label{tab:numerical_data}
\end{table}

\bibliography{references}

\end{document}